%% file: main.tex
\documentclass[sigconf]{acmart}

\AtBeginDocument{%
  \providecommand\BibTeX{{%
    \normalfont B\kern-0.5em{\scshape i\kern-0.25em b}\kern-0.8em\TeX}}}

\setcopyright{acmcopyright}
\copyrightyear{2022}
\acmYear{2022}
\acmDOI{10.1145/3524844.3528051}
\acmConference[SEAMS 2022]{The 17th Symposium on Software Engineering for Adaptive and Self-Managing Systems}{May 21–29, 2022}{Pittsburgh, PA, USA}

\usepackage{algorithm} 
\usepackage{algpseudocode} 
\usepackage{makecell} 
\usepackage{xspace} 
\usepackage[nocomma]{optidef}
\usepackage{enumitem}
\usepackage{graphicx}
\usepackage{subcaption}
\usepackage{multirow}
\usepackage{balance}
\copyrightyear{2022}
\acmYear{2022}
\setcopyright{acmcopyright}\acmConference[SEAMS '22]{17th International Symposium on Software Engineering for Adaptive and Self-Managing Systems}{May 18--23, 2022}{PITTSBURGH, PA, USA}
\acmBooktitle{17th International Symposium on Software Engineering for Adaptive and Self-Managing Systems (SEAMS '22), May 18--23, 2022, PITTSBURGH, PA, USA}
\acmPrice{15.00}
\acmDOI{10.1145/3524844.3528051}
\acmISBN{978-1-4503-9305-8/22/05}
\begin{document}

\title[\approach: Network- and GPU-aware Management of Serverless Functions at the Edge]{\approach: Network- and GPU-aware Management \\of Serverless Functions at the Edge}

\author{Luciano Baresi, Davide Yi Xian Hu, Giovanni Quattrocchi, Luca Terracciano}

\affiliation{%
  \institution{Dipartimento di Elettronica, Informazione e Bioingegneria, Politecnico di Milano}
  \city{Milan}
  \country{Italy}}
\email{{name.surname}@polimi.it}







\renewcommand{\shortauthors}{Luciano Baresi, Davide Yi Xian Hu, Giovanni Quattrocchi, and Luca Terracciano}
\newlist{questions}{enumerate}{2}
\setlist[questions,1]{label*=\textbf{RQ\arabic*},ref=RQ\arabic*}
\setlist[questions,2]{label=(\alph*),ref=\thequestionsi(\alph*)}

\newcommand{\luc}[1]{\textcolor{red}{\textbf{[LB] #1}}}
\newcommand{\gio}[1]{\textcolor{orange}{\textbf{[GQ] #1}}}
\newcommand{\dav}[1]{\textcolor{cyan}{\textbf{[DH] #1}}}
\newcommand{\lt}[1]{\textcolor{green}{\textbf{[LT] #1}}}

\newcommand{\org}[1]{\textcolor{orange}{#1}}
\newcommand{\blue}[1]{\textcolor{blue}{#1}}
\newcommand\approach{\textit{NEPTUNE}\xspace}


\begin{abstract}
Nowadays a wide range of applications is constrained by low-latency requirements that cloud infrastructures cannot meet. Multi-access Edge Computing (MEC) has been proposed as the reference architecture for executing applications closer to users and reducing latency, but new challenges arise: edge nodes are resource-constrained, the workload can vary significantly since users are nomadic, and task complexity is increasing (e.g., machine learning inference). To overcome these problems, the paper presents \approach, a serverless-based framework for managing complex MEC solutions. \approach i) places functions on edge nodes according to user locations, ii) avoids the saturation of single nodes, iii) exploits GPUs when available, and iv) allocates resources (CPU cores) dynamically to meet foreseen execution times. A prototype, built on top of K3S, 
was used to evaluate \approach on a set of experiments that demonstrate a significant reduction in terms of response time, network overhead, and resource consumption compared to three well-known approaches. 
\end{abstract}



\maketitle


\input{sections/introduction}
\input{sections/problem-description}

\input{sections/proposed-solution}

\input{sections/evaluation}
\input{sections/related-works}

\input{sections/conclusions-future-work}
\input{sections/acknowledgements}
\balance

\bibliographystyle{ACM-Reference-Format}
\interlinepenalty=10000
\bibliography{acmart}

\end{document}

%% file: sections/introduction.tex
\section{Introduction}
\label{sec-intro}

Multi-access Edge Computing (MEC) \cite{DBLP:journals/access/PhamFHPLLHD20} has emerged as a new distributed architecture for running computations at the edge of the network and reduce latency compared to cloud executions. 
Differently from cloud computing, which is characterized by a virtually infinite amount of resources placed on large data centers, MEC infrastructures are based on geo-distributed networks of resource-constrained nodes (e.g., 5G base stations) that serve requests and process data close to the users. 

The rise of edge computing~\cite{DBLP:journals/access/GuptaJ15}, also fostered by the advent of 5G networks, enables the creation of applications with extremely low latency requirements like autonomous driving ~\cite{DBLP:journals/pieee/LiuLTYWS19}, VR/AR ~\cite{DBLP:conf/conext/ChoSMMR16} and mobile gaming ~\cite{DBLP:journals/wc/ZhangCZMXHYW19} systems. According to Li et al.~\cite{DBLP:conf/imc/LiYKZ10}, the average network delay from 260 locations to the nearest Amazon EC2 availability zone is approximately 74ms. This makes meeting tight response time requirements in the cloud nearly impossible. In use-cases like obstacle detection, response times of a few hundreds of milliseconds are required~\cite{DBLP:conf/asplos/LinZHSHTM18} and thus the network delay must be lower than the one offered by cloud-based solutions.
Many mobile devices would allow these computations to be executed on the device itself, but this is not always possible given the inherent complexity of some tasks (e.g., machine learning-based ones) and the need for limiting resource consumption (e.g., to avoid battery draining). 

An important challenge of edge computing is that clients usually produce highly dynamic workloads since they move among different areas (e.g., self-driving vehicles) and the amount of traffic in a given region can rapidly escalate (e.g., users moving towards a stadium for an event). To tackle these cases, solutions that scale resources (i.e., virtual machines and containers~\cite{DBLP:conf/ispass/FelterFRR15}) automatically according to the workload  have been extensively investigated in the context of cloud computing, ranging from approaches based on rules~\cite{DBLP:conf/IEEEcloud/DutreilhRMMT10,DBLP:conf/IEEEcloud/YazdanovF14} and machine learning \cite{DBLP:journals/jnca/LiuLSCCC17,DBLP:journals/tsc/ZhuA12} to those based on time-series analysis \cite{DBLP:conf/eurosys/RzadcaFSZBKNSWH20}.
These solutions assume that (new) resources are always available and that nodes are connected through a low-latency, high-bandwidth network. At the edge, these assumptions are not valid anymore and some ad-hoc solutions have been presented in the literature.
For example, Ascigil et al.~ \cite{9326369} and Wang et al.~\cite{DBLP:conf/hpdc/WangAS21} propose a solution for service placement on resource-constrained nodes, while Poularakis et al.~\cite{DBLP:conf/infocom/PoularakisLTTT19} focus on request routing and load balancing at the edge. 

Approaches that focus on service placement or request routing for MEC aim to maximize the throughput of edge nodes, but comprehensive solutions that address placement, routing, and minimal delays at the same time are still work in progress.
In addition, the tasks at the edge, like AI-based computations, are becoming heavier and heavier. The use of GPUs~\cite{DBLP:conf/iccS/CampmanySEMVL16} can be fundamental to accelerate these computations, but they are seldom taken into account explicitly~\cite{DBLP:conf/kivs/KalmbachBKPJH19,DBLP:conf/IEEEcloud/SubediHKR21,DBLP:conf/icsa/BaresiQ20}. 

To tame all these problems, this paper presents \approach, a comprehensive framework for the runtime management of large-scale edge applications that 
exploits placement, routing, network delays, and CPU/GPU interplay in a coordinated way to allow for the concurrent execution of edge applications that meet user-set response times.
\approach uses the \textit{serverless} paradigm ~\cite{DBLP:journals/corr/abs-1902-03383} to let service providers deploy and execute latency-constrained \textit{functions} without managing the underlying infrastructure. \approach uses Mixed Integer Programming (MIP) to allocate functions on nodes, minimize network delays, and exploit GPUs if available. It uses lightweight control theoretical planners to allocate CPU cores dynamically to execute the remaining functions.

\approach is then \textit{holistic} and \textit{self-adaptive}. It addresses edge nodes, function placement, request routing, available resources, and hardware accelerators in a single and coherent framework. Users only provide functions and foreseen response times, and then the system automatically probes available nodes as well as the locality and intensity of workloads and reacts autonomously. While other approaches (see Section~\ref{sec:related}) only focus on a single or few aspects and they can only be considered partial solutions, \approach tackles all of them and oversees the whole lifecycle from deployment to runtime management.


We built a prototype that extends K3S\footnote{https://k3s.io}, a popular tool for container orchestration at the edge, to evaluate \approach on a set of experiments executed on a geo-distributed cluster of virtual machines provisioned on Amazon Web Services (AWS). To provide a comprehensive and meaningful set of experiments, we assessed \approach against three realistic benchmark applications ---including one that can be accelerated with GPUs. The comparison revealed 9.4 times fewer response time violations, and 1.6 and 17.8 times improvements as for resource consumption and network delays, respectively. 

\approach builds up from PAPS~\cite{DBLP:conf/icsoc/BaresiMQ19} from which inherits part of the design. Compared to PAPS, this paper provides i) a new placement and routing algorithm that exploits resources more efficiently and minimizes service disruption, ii) the support of GPUs, and iii) an empirical evaluation of the approach (PAPS was only simulated).

The rest of the paper is organized as follows. Section~\ref{sec:solution} explains the problem addressed by \approach and provides an overview of the solution. Section~\ref{sec:second-level} presents how \approach tackles placement, routing, and GPU/CPU allocation. 
Section~\ref{sec:evaluation} shows the assessment we carried out to evaluate \approach. Section~\ref{sec:related} presents some related work, and Section~\ref{sec:conclusions} concludes the paper.

%% file: sections/problem-description.tex
\newcommand\clusterlevel{Topology level\xspace}
\newcommand\communitylevel{Community level\xspace}
\newcommand\nodelevel{Node level\xspace}
\newcommand\communities{\textit{communities}\xspace}
\newcommand\community{\textit{community}\xspace}
\newcommand\qetime{$QE$\xspace}
\def\qemath{QE}

\section{\approach}
\label{sec:solution}

The goal of \approach is to allow for the execution of multiple concurrent applications on a MEC infrastructure with user-set response times and to optimize the use of available resources. \approach
must be able to take into consideration the main aspects of edge computing: resource-constrained nodes, limited network infrastructure, highly fluctuating workloads, and strict latency requirements.

A MEC infrastructure is composed of a set $N$ of distributed nodes that allow clients to execute a set $A$ of applications.
Edge clients (e.g., self-driving cars, mobile phones, or VR/AR devices) access applications placed on MEC nodes. Each node comes with cores, memory, and maybe GPUs, along with their memory. Because of user mobility, the workload on each node can vary frequently, and resource limitations do not always allow each node to serve all the requests it receives; some requests must be outsourced to nearby nodes. 

\begin{figure}[t]
    \centering
    \includegraphics[width=0.9\linewidth]{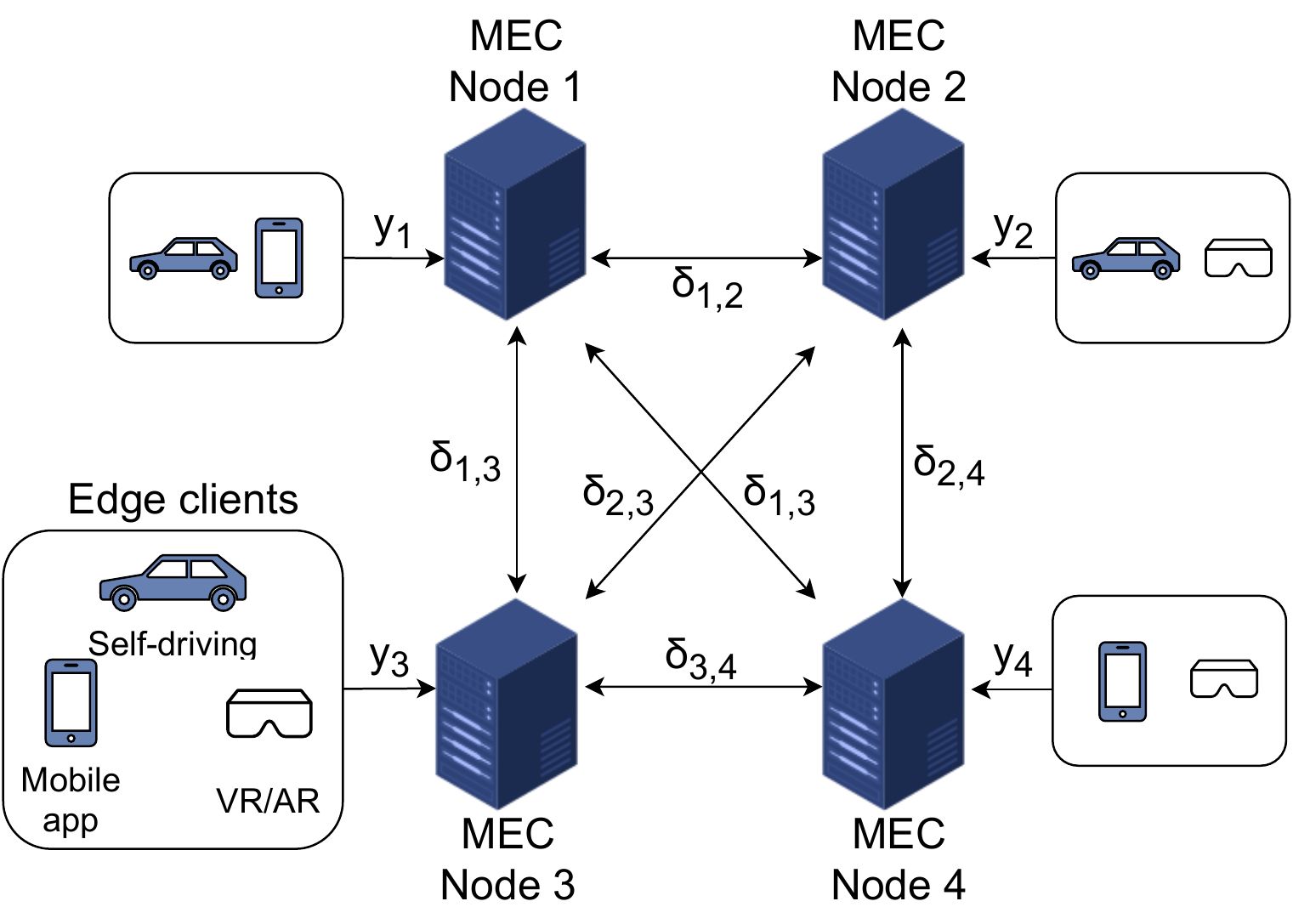}
    \caption{MEC Topology.}
    \label{fig:solution:mec_architecture}
    \Description{}
\end{figure}

Given a request $r$ for an application $a \in A$, its response time $RT$ is measured as the time required to transmit $r$ from the client to the closest node $i$, execute the request, and receive a response.
More formally, $RT$ is defined as $RT=E+Q+D$, where $E$ (\textit{execution time}) represents the time taken for running $a$, $Q$ (\textit{queue time}) is the time spent by $r$ waiting for being managed, and $D$ is the network delay (or network latency). In particular, as shown in Figure \ref{fig:solution:mec_architecture}, $D$ is the sum of the (round trip) time needed by $r$ to reach $i$ ($y_i$) and, if needed, the (round trip) time needed to outsource the computation on a nearby node $j$ ($\delta_{i,j}$). \approach handles requests once they enter the MEC topology and assumes $y_i$ be optimized by existing protocols \cite{DBLP:journals/rfc/rfc3626}.

\approach is then \textit{location aware} since it considers the geographical distribution of both nodes and workloads. It stores a representation of the MEC network by measuring the inter-node delays and it monitors where and how many requests are generated by users. Furthermore, \approach allows users to put a threshold (service level agreement) on the response time provided by each application ($RT^R_a$).  

\subsection{Solution overview}
\label{sec:solution:overview}

\approach requires that an application $a$ be deployed as a set $F_a$ of functions ---as prescribed by the serverless paradigm. Developers focus on application code without the burden of managing the underlying infrastructure. Each function covers a single functionality and supplies a single or a small set of REST endpoints. The result is more flexible and faster to scale compared to traditional architectures (e.g., monoliths or microservices).

Besides the function's code, \approach requires that each function be associated with a user-provided required response time $RT^R_f$ (with $RT^R_f \leq RT^R_a$) and the memory $m^{CPU}_f$ needed to properly execute it. In case of GPU-accelerated functions, the GPU memory $m^{GPU}_{f}$ must also be specified. Then, \approach manages both its deployment by provisioning one or more function \textit{instances}, and its operation with the goal of fulfilling the set response time.

\approach manages functions through a three-level control hierarchy: \textit{Topology}, \textit{Community}, and \textit{Node}.

Since function placement is NP-hard~\cite{DBLP:conf/infocom/PoularakisLTTT19}, the main goal of the \clusterlevel is to tackle the complexity for the lower levels by splitting the topology into \communities of closely-located nodes. Each \community is independent of the others and a request can only be managed within the community of the node that received it. If this is not possible, then the community is undersized and the \clusterlevel must reconfigure the communities.

The \clusterlevel employs a single controller based on the \textit{Speaker-listener Label Propagation Algorithm} (SLPA) --- proposed by Xie et al. ~\cite{DBLP:conf/icdm/XieSL11} --- to create the communities. 
SLPA has a complexity of $O(t*N)$ where $N$ is the number of distributed nodes and $t$ is the user-defined maximum number of iterations. Since the complexity scales linearly with the number of nodes, this solution has proven to be suitable also for large clusters~\cite{DBLP:conf/icsoc/BaresiMQ19}.

Given a maximum community size $MCS$ and the maximum allowed network delay $\Delta$, SLPA splits the topology in a set of communities with a number of nodes that is lower than $MCS$ and inter-node delays smaller than $\Delta$ (i.e., $\delta_{i,j}\le \Delta$  for all nodes $i$ and $j$ in the community).
SLPA could potentially assign a node to multiple communities, but to avoid resource contention, \approach re-allocates shared nodes ---if needed--- to create non-overlapping communities. 

At \communitylevel, each community is equipped with a MIP-based controller in charge of managing function instances. The controller places function instances on the different nodes of the community by first considering those that could exploit GPUs, if available. The goal is to minimize network delay by dynamically deploying function instances close to where the demand is generated and at the same time to minimize the time spent to forward them when needed.

The \communitylevel computes routing policies i) to allow each node to forward part of the workload to other close nodes, and ii) to prioritize computation-intensive functions by forwarding requests to GPUs up to their full utilization, and then send the remaining requests to CPUs.
To avoid saturating single nodes, the \communitylevel can also scale function instances horizontally, that is, it can replicate them on nearby nodes. 
For example, if a node $i$ cannot offer enough resources to execute an instance of $f$ to serve workload $\lambda_{f,i}$, the \communitylevel creates a new instance (of $f$) on a node $j$ close to $i$, and the requests that cannot be served by $i$ are forwarded to $j$.

\begin{figure}[t]
    \centering
    \includegraphics[width=\linewidth]{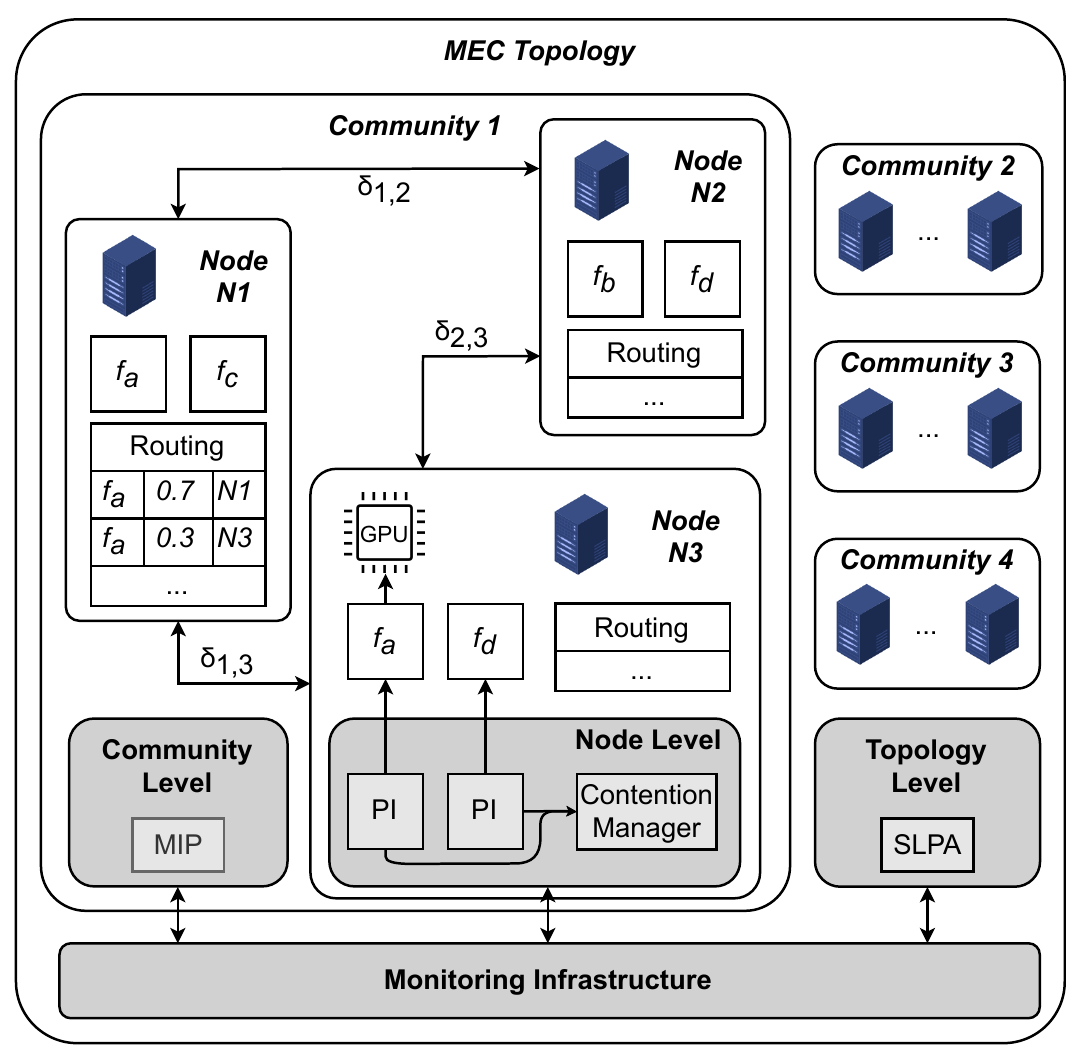}
    \caption{\approach.}
    \label{fig:solution:control-level}
    \Description{\approach.}
\end{figure}

While the first two control levels take care of network latency ($D$), once the requests arrive at the node that processes them, the \nodelevel ensures that function instances have the needed amount of cores to meet set response times. Each function instance is managed by a dedicated \textit{Proportional Integral} (PI) controller that provides vertical scaling, that is, it adds, or removes, CPU cores to the function (container). Unlike other approches~\cite{DBLP:conf/IEEEcloud/RattihalliGLT19, DBLP:conf/noms/BallaSM20}, \approach can reconfigure CPU cores without restarting function instances, that is, without service disruption. Figure \ref{fig:solution:control-level} shows that, given a set point (i.e., the desired response time), the PI controller periodically monitors the performance of its function instance and dynamically allocates CPU cores to optimize both execution time $E$ and queue time $Q$. 

Also, the controllers at \nodelevel are independent of each other: each controller is not aware of how many cores the others would like to allocate. Therefore, since the sum of the requests might exceed the capacity of a node, a \textit{Contention Manager} solves resource contentions by proportionally scaling down requested allocations. 

Every control level is meant to work independently of the others, and no communication is required between controllers operating at the same level.
This means that \approach can easily \textit{scale}.

The three control levels operate at different frequencies but yet cohesively to eliminate potential interference. Thanks to fast PI controllers and vertical scaling, the \nodelevel operates every few seconds to handle workload bursts, whereas the \communitylevel computes functions and routing policies in the order of minutes, and allows the \nodelevel to fully exploit the underlying resources. The \clusterlevel runs at longer intervals, but it can react faster when there are changes in the network, such as when a node is added or removed.

A real-time monitoring infrastructure is the only communication across the levels. It allows \approach to gather performance metrics (e.g., response times, core allocations per function instance, network delays) needed by the three control levels to properly operate. Note that the controllers at Community and Node levels measure device performance in real-time without any a-priori assumption. This means that \approach can also manage heterogeneous CPUs with different performance levels (e.g., different types of virtual machines).

Figure~\ref{fig:solution:control-level} shows a MEC topology controlled by the three-level hierarchical control adopted by \approach. Control components are depicted in grey.
First of all, the \clusterlevel splits the network into four communities.
Each community is handled by its \communitylevel controller, responsible for function placement and request routing.
The figure reports a detailed view of Community 1.
We can observe that 
i) a set of functions ($f_a$, $f_b$, $f_c$, $f_d$) is placed across three nodes and,
ii) each node is provided with its set of routing policies.
In particular, the routing policies for Node N1 enforce that 70\% of the requests for $f_a$ are served by the node itself, while the remaining 30\% is forwarded to the instance, running on Node N3, that exploits a GPU.
Finally, the figure shows the materialization of \nodelevel on Node N3.
Each function instance ($f_a$, $f_d$) is vertically scaled by a dedicated PI controller, and resource contentions are solved by the \textit{Contention Manager}.

%% file: sections/proposed-solution.tex
\section{Placement,  Routing, and Allocation}\label{sec:second-level}
As explain above, the \clusterlevel partitions the network in a set of communities using algorithm SLPA. Each community is controlled independently from the others.
The main goal of the \communitylevel controller is to dynamically place function executions as close to where the workload is generated as possible. 
A trivial solution, which can drastically reduce network delay, would be to replicate the whole set of functions on each node. Since nodes have limited resources, this approach is often not feasible.

An effective placement solution should not saturate nodes and should allow for stable placement, that is, it should avoid the disruption implied by keeping migrating functions among nodes\footnote{\approach does not handle application state migration.}. At the same time, the placement cannot be fully static since users move and requirements change.

Each placement effort should also consider \textit{graceful termination periods} and \textit{cold starts}. The former is the user-defined amount of time we must wait whenever a function instance is to be deleted to let it finish serving pending requests.
The latter is a delay that can range from seconds to minutes and that affects newly created instances~\cite{DBLP:conf/usenix/WangLZRS18}. Before a function instance starts serving requests, a container must be started and the execution environment be initialized. Diverse approaches, like container pooling~\cite{DBLP:journals/corr/abs-1903-12221,DBLP:conf/middleware/SilvaFP20}, can mitigate cold starts, but
their efficiency depends on the functions at hand and they cannot always reduce cold starts in a significant way. This is why \approach does not exploit these solutions natively. 

To address these problems, the \communitylevel adopts two similar instances of a 2-step optimization process ---based on Mixed Integer Programming--- to allocate GPUs first and then CPUs. In both cases, 
the first step aims to find the best function placement and routing policy that minimize the overall network delay. 
Then, since diverse placements with a network delay close to the optimal one may require different changes in function deployment, the second step is in charge of choosing a placement that minimizes deployment actions, that is, disruption.

Table~\ref{tab:MIP-input} summarizes 
the inputs \approach requires to the user, the characteristics of used nodes, the values gathered by the monitoring infrastructure, and the decision variables adopted in the MIP formulation.

\begin{table}
    \small
    \caption{Inputs, data, and decision variables.}
    \label{tab:MIP-input}
    \begin{tabular}{cl}
        \toprule
        \textbf{Inputs} & \\
        \midrule
        $m^{CPU}_f$      & Memory required by function $f$                \\
        $m^{GPU}_f$      & GPU memory required by function $f$            \\
        $\phi_f$         & Maximum allowed network delay for function $f$  \\
        \midrule
        \multicolumn{2}{l}{\textbf{Infrastructure data}} \\
        \midrule
        $M^{CPU}_j$      & Memory available on node $j$ \\
        $M^{GPU}_j$      & GPU memory available on node $j$ \\
        $U^{CPU}_j$            & CPU cores on node $j$ \\
        $U^{GPU}_j$            & GPU cores on node $j$ \\
        \midrule
        \multicolumn{2}{l}{\textbf{Monitored data}} \\
        \midrule
        $\delta_{i,j}$   & Network delay between nodes $i$ and $j$ \\
        $O_{best}$       & Objective function value found after step 1  \\
        $\lambda_{f,i}$  & Incoming $f$ requests to node $i$ \\
        $u^{CPU}_j$            & Average CPU cores used by node $j$ per single $f$ request \\
        $u^{GPU}_j$            & Average GPU cores used by node $j$ per single $f$ request \\
        \midrule
        \multicolumn{2}{l}{\textbf{Decision variables}} \\
        \midrule
        $x^{CPU}_{f,i,j}$      & Fraction of $f$ requests sent to CPU instances from node $i$ to $j$ \\
        $c^{CPU}_{f,j}$        & $1$ if a CPU instance of $f$ is deployed on node $j$, $0$ otherwise\\
        $x^{GPU}_{f,i,j}$      & Fraction of $f$ requests sent to GPU instances from node $i$ to $j$ \\
        $c^{GPU}_{f,j}$        & $1$ if a GPU instance of $f$ is deployed on node $j$, $0$ otherwise\\
        $MG_f$           & Number of $f$ migrations                     \\
        $CR_f$           & Number of $f$ creations                     \\
        $DL_f$           & Number of $f$ deletions                    \\
        \bottomrule
    \end{tabular}
\end{table}



\subsection{Function placement}
Each time the \communitylevel is activated, the 2-step optimization process is executed twice. The first execution aims to fully utilize GPU resources, while the second only considers CPUs and the remaining workload to be handled.

Since the two executions are similar, the formulation presented herein is generalized.
Some of the employed data are resource-specific (e.g., $x_{f,i,j}$, $m_{f}$): Table~\ref{tab:MIP-input} differentiate them with a $CPU$ or $GPU$ superscript, while in the rest of this section the superscripts are omitted for simplicity.

\label{sec:second-level:first-step}
\paragraph{Network delay minimization.}
The first step aims to place function instances and to find routing policies that minimize the overall network delay $D$ in a given community $C \subseteq N$.

The formulation employs two decision variables: $x_{f,i,j}$ and $c_{f,j}$.
The former ($x_{f,i,j} \in [0:1]$) represents the amount of incoming $f$ requests\footnote{For the sake of brevity, we use $f$ request to mean a request generated for function $f$, and $f$ instance to refer to an instance of function $f$.} ($\lambda_{f,i}$) that node $i$ forwards to node $j$ (i.e., routing policies). The latter ($c_{f,j}$) is a 
boolean variable that is $true$ if an $f$ instance is deployed onto node $j$ (i.e., placement). 

The objective function (Formula \ref{eq:first-step-cpu-objective}) minimizes the overall network delay of the incoming workload in $C$. Starting from the incoming $f$ requests to each node $i$ and the measured delay between nodes $i$ and $j$, it computes the fractions of outsourced requests to minimize the overall network delay:

\begin{equation}
    \begin{aligned}
        \label{eq:first-step-cpu-objective}
         & min \quad \sum_{f}^{F}\sum_{i}^{C}\sum_{j}^{C}x_{f,i,j} * \lambda_{f,i} * \delta_{i,j}
    \end{aligned}
\end{equation}

If we only considered inter-node delays ($\delta_{i,j}$), we would minimize the overall network delay only if incoming requests were distributed evenly (i.e., each node manages the same amount of requests). Since workloads can be very different, the addition of per-node incoming requests ($\lambda_{f,i}$) gives a more appropriate formulation of the problem. Intuitively, the higher the workload in a specific area is, the more important the minimization of network delay becomes.

In addition to the function to minimize, we must add some constraints.
First, requests cannot be forwarded too far from where they are generated. Each function is characterized by parameter $\phi_f$, which sets the maximum allowed network delay of each $f$ request:

\begin{equation}
    \begin{aligned}
        \label{eq:first-step-cpu-constraint-SLA-violation}
         & x_{f,i,j} * \delta_{i,j}\leq x_{f,i,j} * \phi_f \quad \forall i,j \in C, \forall f \in F
    \end{aligned}
\end{equation}

Second, the nodes that receive forwarded requests must have a function instance that can serve them:

\begin{equation}
    \begin{aligned}
        \label{eq:first-step-cpu-constraint-replica-existence}
         & c_{f,j}= if \ (\sum_{i}^{C} x_{f,i,j} > 0) \ 1 \ else~0 \quad \forall j \in C, \forall f \in F
    \end{aligned}
\end{equation}

Third, the overall memory required by the functions ($m_f$) placed on a node must not exceed its capacity $M_j$:

\begin{equation}
    \begin{aligned}
        \label{eq:first-step-resource-memory}
         & \sum_{f}^{F} c_{f,j} * m_f\leq M_j \quad \forall j \in C \\
    \end{aligned}
\end{equation}

\noindent

Fourth, to avoid resource contentions, routing policies must consider the overall amount of GPU or CPU cores available in the node ($U_j$) and the average GPU or CPU cores consumption for each $f$ request processed on node $j$ ($u_{f,j}$):
\begin{equation}
    \begin{aligned}
        \label{eq:first-step-resource-cpu}
         & \sum_{i}^{C} \sum_{f}^{F} x_{f,i,j} * \lambda_{f,i} * u_{f,j}\leq U_j \quad \forall j \in C \\
    \end{aligned}
\end{equation}
\noindent

Fifth, routing policies must be defined for all the nodes in the community and all the functions of interest:

\begin{equation}
    \begin{aligned}
        \label{eq:first-step-cpu-constraint-process-all-requests}
         & \sum_{j}^{C} x_{f,i,j}= 1 \quad \forall i \in C, \forall f \in F
    \end{aligned}
\end{equation}

Note that when $i = j$, $x_{f,i,j}$ gives the fraction of $f$ requests executed locally, that is, on $i$ itself. 

This optimization problem finds the best placement with the minimum network delay. However, each iteration (execution of the optimization problem) may suggest a placement that requires many disruptive operations (i.e., deletions, creations, and migrations) with respect to the previous placement (iteration). For this reason, a second step is used to minimize service disruption and ameliorate the result.

\paragraph{Disruption minimization}
\label{sec:second-level:second-step}

The second step searches for a function placement that minimizes function creation, deletion, and migration with an overall network delay close to the optimal one found by the first step.

This means that the second step keeps the constraints defined in Formulae~\ref{eq:first-step-cpu-constraint-SLA-violation}-~\ref{eq:first-step-cpu-constraint-process-all-requests} and adds: 

\begin{equation}
    \begin{aligned}
        \label{eq:second-step-constraint-delay-neigh}
         \sum_{f}^{F}\sum_{i}^{C}\sum_{j}^{C}x_{f,i,j} * \lambda_{f,i} * \delta_{i,j} \leq O_{best} * (1+\epsilon)
    \end{aligned}
\end{equation}

to impose that the final placement must be in the interval $[O_{best},$ $O_{best} * (1+\epsilon)]$,
where $O_{best}$ is the smallest network delay found after the first step, and $\epsilon$ is an arbitrarily small parameter that quantifies the worsening in terms of network overhead. For example, $\epsilon = 0.05$ means a worsening up to $5\%$. 

We also consider the number of created, deleted, and migrated $f$ instances between two subsequent executions of the 2-step optimization process, that is, between the to-be-computed placement ($c_{f,i}$) and the current one ($c^{old}_{f,i}$).
$DL_f$ and $CR_f$ denote, respectively, the maximum between $0$ and the removed (added) $f$ instances between the two iterations:


    \begin{equation}
        \begin{aligned}
            \label{eq:second-step-formula-creations-deletions}
             & DL_f=\sum_{i}^{C}max(c^{old}_{f,i} - c_{f,i}, 0) \quad \forall f \in F\\
             & CR_f=\sum_{i}^{C}max(c_{f,i} - c^{old}_{f,i}, 0) \quad \forall f \in F\\
        \end{aligned}
    \end{equation}

The number of migrations (in the new placement) that represents the number of instances that have been moved from one node to another is computed as the minimum between instance creations $CR_f$ and instance deletions $DL_f$:


    \begin{equation}
        \begin{aligned}
        \label{eq:second-step-formula-migration}
         & MG_f=min(CR_f, DL_f) \quad \forall f \in F\\
        \end{aligned}
    \end{equation}





The new objective function is then defined as:

\begin{equation}
    \begin{aligned}
        \label{eq:second-step-objective}
         & min \quad \sum_{f}^{F}MG_f + \frac{1}{DL_f + 2} - \frac{1}{CR_f + 2}
    \end{aligned}
\end{equation}

The goal of the objective function is to minimize the number of migrations ($MG_f$), since deletions and creations are necessary, to avoid over- and under-provisioning. Factors $\frac{1}{DL_f + 2}$ and $\frac{1}{CR_f + 2}$, which are always lower than 1, allow us to discriminate among solutions with the same amount of migrations, but a different number of creations and deletions.

This formulation ensures close-to-optimal network delays, along with the minimum number of instances, to serve the current workload. The controllers at \nodelevel are then entitled to scale executors vertically as needed (see next Section).

\label{sec:second-level:heterogeneous-hw}

\subsection{CPU allocation}\label{sec:first-level}

The \nodelevel is in charge of minimizing \qetime, that is, the handling time, defined as the sum of the execution time $E$ and the queue time $Q$; the network delay $D$ is already minimized by the \communitylevel. \qetime can vary due to many factors, such as variations in the workload or changes in the execution environment and we aim to control it by changing the amount of CPU cores allocated to function instances. If this is not enough, the problem is lifted up to the \communitylevel that re-calibrates the number of function instances.

Control theoretical approaches have proven to be an effective solution for the self-adaptive management of these resources~ \cite{DBLP:conf/icsa/BaresiQ20,DBLP:conf/globecom/GrimaldiPPSS15}.
The \nodelevel comprises a lightweight \textit{Proportional-Integral} (PI) controller for each function instance to scale allocated cores dynamically.
PI controllers support fast control periods, have constant complexity, and provide formal design-time guarantees.

Each function instance is equipped with an independent PI controller. The control loop monitors the average value of \qetime, computes the allocation, and actuates it.
More formally, given a desired set point $\qemath_{f,desired}$, the controller periodically measures the current value of $\qemath_{f,j}$ (controlled variable) --- the actual value of $\qemath_f$ on node $j$ --- and computes the delta between desired and actual value.
Note that, since the controllers will strive to keep $\qemath_{f,j}$ close to the set point $\qemath_{f,desired}$, this value should be set to a lower value than the desired $RT^R_f$. 

The controller reacts to the error and recommends the new amount of cores that the function should use. Algorithm \blue{\ref{alg:computecores}} describes the computation:
\begin{algorithm}
    \caption{\nodelevel CPU core allocation.}
    \begin{algorithmic}[1]
        \Procedure{ComputeInstanceCores}{$f$, $j$}
        \State $err := \frac{1}{\qemath_{f,desired}} - \frac{1}{\qemath_{f,j}}$;
        \State $cpu := getCPUAllocation(f,j)$;
        \State $int^{old} := cpu - g_{int} * err^{old}$;
        \State $int := int^{old} + g_{int} * err$;
        \State $err^{old} := err$;
        \State $prop := err * g_{prop}$;
        \State $cpu := int + prop$;
        \State $cpu := max(cpu_{min}, min (cpu_{max}, cpu))$;
        \EndProcedure
    \end{algorithmic}
    \label{alg:computecores}
\end{algorithm}
Line 2 computes error $err$ as the difference between the inverse of $\qemath_{f,desired}$ and $\qemath_{f,j}$.
To compute the Integral contribution, the current core allocation ($cpu$) of the function instance is retrieved at line 3. The previous integral contribution $int^{old}$ is computed at line 4 by using the allocation, the integral gain $g_{int}$ (i.e., a tuning parameter), and the prior error $err^{old}$. 
The integral component $int$ is computed by multiplying the current error $err$ times the integral gain $g_{int}$, and by adding $int^{old}$ (line 5). The previous error $err^{old}$ is then updated at line 6.

The proportional contribution is computed by using $err$ and the proportional gain $g_{prop}$ at line 7.
Finally, the new allocation is calculated as the sum of the two contributions (line 8) and then adjusted according to the maximum and minimum allowed core allocations $cpu_{max}$ and $cpu_{min}$, respectively.

Being independent of the others, these controllers 
are not aware of available CPU cores and of the allocations computed by the other controllers.
Therefore, the computed allocations (line 9) are not immediately applied since they could exceed the allowed capacity. The allocations of the function instances deployed on a node are processed by a \textit{Contention Manager} (one per node), which is in charge of computing a feasible allocation. If the sum of suggested allocations fits the allowed capacity, they are applied without any modification. Otherwise, they are scaled down proportionally.
%
%
The \textit{Contention Manager} can easily be extended and embed other, non-proportional heuristics to manage resource contention.

%% file: sections/evaluation.tex
\section{evaluation}\label{sec:evaluation}






\paragraph{Implementation}
We implemented a prototype\footnote{Source code available at \url{https://github.com/deib-polimi/edge-autoscaler}.} of \approach built on top K3S, a popular distribution of Kubernetes\footnote{\url{https://kubernetes.io/}.} optimized for edge computing.
Each control level is materialized in a dedicated component that exclusively uses native K3s APIs to manage deployed applications. Conversely to existing approaches (see Section \ref{sec:related}), the prototype is capable of performing \textit{in-place} vertical scaling of containers, that is, it can dynamically update the CPU cores allocated to the different containers without restarting the application.

The stable version of K3S does not allow one to change allocated resources without restarting function instances, a process that sometimes can take minutes. This could decrease the capability of the \nodelevel to handle bursty workloads.
For this reason, the prototype augments K3S with the \textit{Kubernetes Enhancement Proposal} 1287 that implements \textit{In-Place Pod Vertical Scaling}\footnote{https://github.com/kubernetes/enhancements/tree/master/keps/sig-node/1287-in-place-update-pod-resources} and allows resources to be changed without restarts. This enables faster control loops and better control quality.

To provide an effective usage of GPUs, the prototype uses \textit{nvidia-docker}\footnote{https://github.com/NVIDIA/nvidia-docker}, a container runtime that enables the use of GPUs within containers.
However, by default, GPU access can only be reserved to one function instance at a time. This prevents the full exploitation of GPUs and limits the possible placements produced at \communitylevel.
To solve this problem, the prototype employs a \textit{device plugin}\footnote{https://github.com/awslabs/aws-virtual-gpu-device-plugin} developed by Amazon that enables the fractional allocation of GPUs. In particular, the plugin makes use of the \textit{Nvidia Multi Process Service}\footnote{https://docs.nvidia.com/deploy/mps/index.html} (MPS), a runtime solution designed to transparently allow GPUs to be shared among multiple processes (e.g., containers).

\paragraph{Research questions}
The solution adopted at \clusterlevel has been largely covered by PAPS~\cite{DBLP:conf/icsoc/BaresiMQ19}. The experiments in the paper focus on evaluating Community and \nodelevel.
The conducted evaluation addresses the following research questions:

\begin{questions}
  \item How does \approach handle workloads generated by mobile users at the edge? 
  \item How does \approach perform compared to other state-of-the-art approaches?
  \item How does \approach use GPUs to speed up response times?
\end{questions}

\subsection{Experimental setup}
\label{sec:evaluation:experiment_setup}

\paragraph{Infrastructure}
We conducted the experiments on a simulated MEC topology with nodes provisioned as a cluster of AWS EC2 geo-distributed virtual machines distributed across three areas. Each area corresponds to a different AWS region: Area A to \textit{eu-west},  Area B to \textit{us-east}, and Area C to \textit{us-west}.
Since communities are independent, our experiments focused on evaluating different aspects of \approach within a single community that included the three areas. 

\begin{figure}[t]
  \centering
  \includegraphics[width=0.9\linewidth]{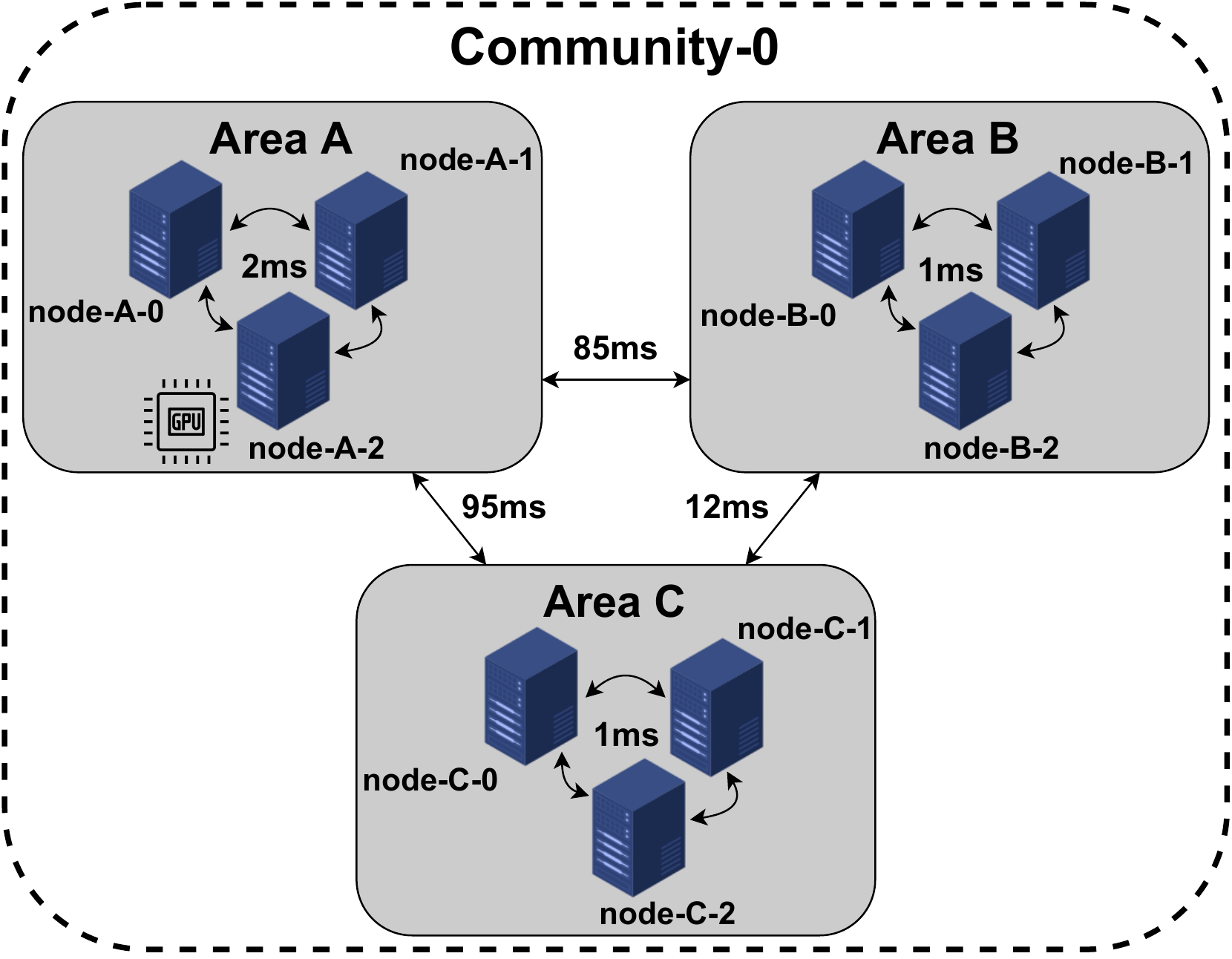}
  \caption{Network delay between areas.}
  \label{fig:evaluation:netdelay}
  \Description{Network delay between areas.}
\end{figure}

Figure \ref{fig:evaluation:netdelay} shows the average network delays between each pair of areas and nodes computed as the round trip times of an ICMP \cite{DBLP:journals/rfc/rfc777} (Internet Control Message Protocol) packet. Note that nodes of the same area were deployed onto different AWS availability zones to obtain significant network delays.
Each area contained three worker nodes, and one in Area A was GPU-empowered.
These nodes were deployed as \textit{c5.xlarge} instances (4 vcpus, 8 GB memory); the one with GPU that used a \textit{g4dn.xlarge} instance (4 vcpus, 16 GB memory, 1 GPU). The master node (not depicted in the figure) was deployed on a \textit{c5.2xlarge} instance (8 vcpus, 16 GB memory).

\paragraph{\approach control periods}
Node controllers were configured with a control period of $5$ seconds. Faster control loops can be used but they may lead to inconsistent resource allocation updates since K3S resource states are stored in a remote database.
Function placement and routing policies were recomputed by Community controllers each $1$ minute while Topology controller was triggered every $10$ minutes.

\paragraph{Applications}
To work on a reasonable set of experiments, we used the three applications summarized in Table~\ref{Tab:functions}: we created the first function, and we borrowed the other two from the literature~\cite{DBLP:conf/closer/QuintK18,DBLP:conf/ccbd/ShiWXC16}. These applications are written using multiple programming languages (e.g., Rust, Java, Go) and have different memory requirements (ranging from 15MB to 500MB) and cold start times (from a bunch of seconds to minutes).
The first application is \textit{primes}, a stateless and CPU-heavy function that counts all the prime numbers less than a given input number.
As exemplar complex application we employed \textit{sock-shop}\footnote{\url{https://github.com/microservices-demo/microservices-demo}} that implements an e-commerce platform.
The application uses a microservice architecture; we further decomposed it into smaller functions to make it suitable for a serverless platform\footnote{The source code of the function-based version of \textit{sock-shop} is available at \url{https://github.com/deib-polimi/serverless-sock-shop}}. For example, microservice \textit{carts} was divided into three smaller units: \textit{carts-post}, \textit{carts-delete} and \textit{cart-util}.
Finally, to also evaluate GPU-accelerated tasks (e.g machine learning inference), we used \textit{Resnet}~\cite{DBLP:conf/uemcom/VermaQF17}, a neural network model for image classification, implemented using TensorFlow Serving.
For each function Table~\ref{Tab:functions} also reports the memory requirements, the cold start times and the desired response times (obtained by applying the procedure described in Section \ref{sec:evaluation:threats}). The set points used by PI controllers were set to half of the value of $RT^R_f$.

We used \textit{Locust}\footnote{\url{https://locust.io/}}, a distributed scalable performance testing tool, to feed the system, and mimicked service demand $\lambda_{f,i}$ through different realistic, dynamic workloads. Each experiment was executed five times to have (more) consistent results.

\begin{table}[t]
  \caption{Characteristics of deployed functions.}
  \begin{tabular}{|lllll|}
    \hline
    \multicolumn{1}{|l|}{Name}         & \multicolumn{1}{l|}{Language} & \multicolumn{1}{l|}{Memory}       & \multicolumn{1}{l|}{$RT^R_f$}   & Cold start   \\ \hline
    \multicolumn{5}{|c|}{Simple stateless function}                                                                                                    \\ \hline
    \multicolumn{1}{|l|}{primes}       & \multicolumn{1}{l|}{Rust}     & \multicolumn{1}{l|}{$\sim$15 MB}  & \multicolumn{1}{l|}{200ms} & \textless 5s \\ \hline
    \multicolumn{5}{|c|}{Complex application}                                                                                                          \\ \hline
    \multicolumn{1}{|l|}{carts-post}   & \multicolumn{1}{l|}{Java}     & \multicolumn{1}{l|}{$\sim$360 MB} & \multicolumn{1}{l|}{300ms} & $\sim$100s   \\ \hline
    \multicolumn{1}{|l|}{carts-delete} & \multicolumn{1}{l|}{Java}     & \multicolumn{1}{l|}{$\sim$360 MB} & \multicolumn{1}{l|}{200ms} & $\sim$100s   \\ \hline
    \multicolumn{1}{|l|}{carts-util}   & \multicolumn{1}{l|}{Java}     & \multicolumn{1}{l|}{$\sim$360 MB} & \multicolumn{1}{l|}{200ms} & $\sim$100s   \\ \hline
    \multicolumn{1}{|l|}{catalogue}    & \multicolumn{1}{l|}{Go}       & \multicolumn{1}{l|}{$\sim$15 MB}  & \multicolumn{1}{l|}{200ms} & \textless 5s \\ \hline
    \multicolumn{1}{|l|}{orders}       & \multicolumn{1}{l|}{Java}     & \multicolumn{1}{l|}{$\sim$400 MB} & \multicolumn{1}{l|}{600ms} & $\sim$100s   \\ \hline
    \multicolumn{1}{|l|}{payment}      & \multicolumn{1}{l|}{Go}       & \multicolumn{1}{l|}{$\sim$15 MB}  & \multicolumn{1}{l|}{50ms}  & \textless 5s \\ \hline
    \multicolumn{1}{|l|}{shipping}     & \multicolumn{1}{l|}{Java}     & \multicolumn{1}{l|}{$\sim$350 MB} & \multicolumn{1}{l|}{50ms}  & $\sim$100s   \\ \hline
    \multicolumn{1}{|l|}{login}        & \multicolumn{1}{l|}{Go}       & \multicolumn{1}{l|}{$\sim$15 MB}  & \multicolumn{1}{l|}{100ms} & \textless 5s \\ \hline
    \multicolumn{1}{|l|}{registration} & \multicolumn{1}{l|}{Go}       & \multicolumn{1}{l|}{$\sim$15 MB}  & \multicolumn{1}{l|}{200ms} & \textless 5s \\ \hline
    \multicolumn{1}{|l|}{user}         & \multicolumn{1}{l|}{Go}       & \multicolumn{1}{l|}{$\sim$15 MB}  & \multicolumn{1}{l|}{50ms}  & \textless 5s \\ \hline
    \multicolumn{5}{|c|}{Machine Learning inference}                                                                                                   \\ \hline
    \multicolumn{1}{|l|}{resnet}       & \multicolumn{1}{l|}{Python}   & \multicolumn{1}{l|}{$\sim$500 MB}  & \multicolumn{1}{l|}{550ms} & $\sim$100s \\ \hline
  \end{tabular}
  \label{Tab:functions}
\end{table}

\paragraph{Collected metrics}
For each experiment, we collected the average ($\mu$) and standard deviation ($\sigma$) of the following metrics: i) \textit{response time} (ms) as defined in Section~\ref{sec:solution}, ii) \textit{response time violation rate} (\% of requests) defined as the percentage of requests that are not served within $RT^R_f$ considering the 99th percentile of the measured response times, iii)
\textit{network time rate} (\%) as the percentage of time spent to forward requests in the network over the total response time ($D/RT$), and iv) \textit{allocated cores} (millicores or thousandths of a core) to measure the resources consumed by function instances.

\paragraph{Competitors} 

Our experiments compare \approach against three well-known approaches: K3S,  \textit{Knative}\footnote{\url{https://knative.dev/docs/}.} (KN), and \textit{OpenFaaS}\footnote{\url{https://www.openfaas.com/}.} (OF).
K3S is one of the most popular solutions for container orchestration at the edge. It manages the full lifecycle of containerized applications deployed in a topology and adopts a fair placement policy, that is, it schedules containers to keep the resource utilization of nodes equal. K3S exploits the \textit{Horizontal Pod Autoscaler}\footnote{\url{https://rancher.com/docs/rancher/v2.5/en/k8s-in-rancher/horitzontal-pod-autoscaler/}.} to horizontal scale applications.
KN and OF add serverless functionalities to K3S and a set of custom components to perform request routing and horizontal scaling.

To achieve consistent and statistical relevant results, all experiments described in this section were run 5 times.

\subsection{RQ1: Moving workload}
\label{sec:evaluation:geo_workload}

The first experiments evaluate the performance of \approach when users move between Area A and Area B within the same community.
We used a cluster of four worker nodes: two nodes in Area A (not equipped with GPU) and two in Area B.
Each run lasted $60$ minutes and used application \textit{primes} with $RT^{R}_{primes}$ set to 200ms and the set point of PI controllers to 100ms. User migration happened twice per run and consisted in moving 100 users from one area to another in less than 10 minutes.


\begin{figure}[t]
  \centering
  \begin{subfigure}{1\columnwidth}
    \centering
    \includegraphics[width=1\columnwidth]{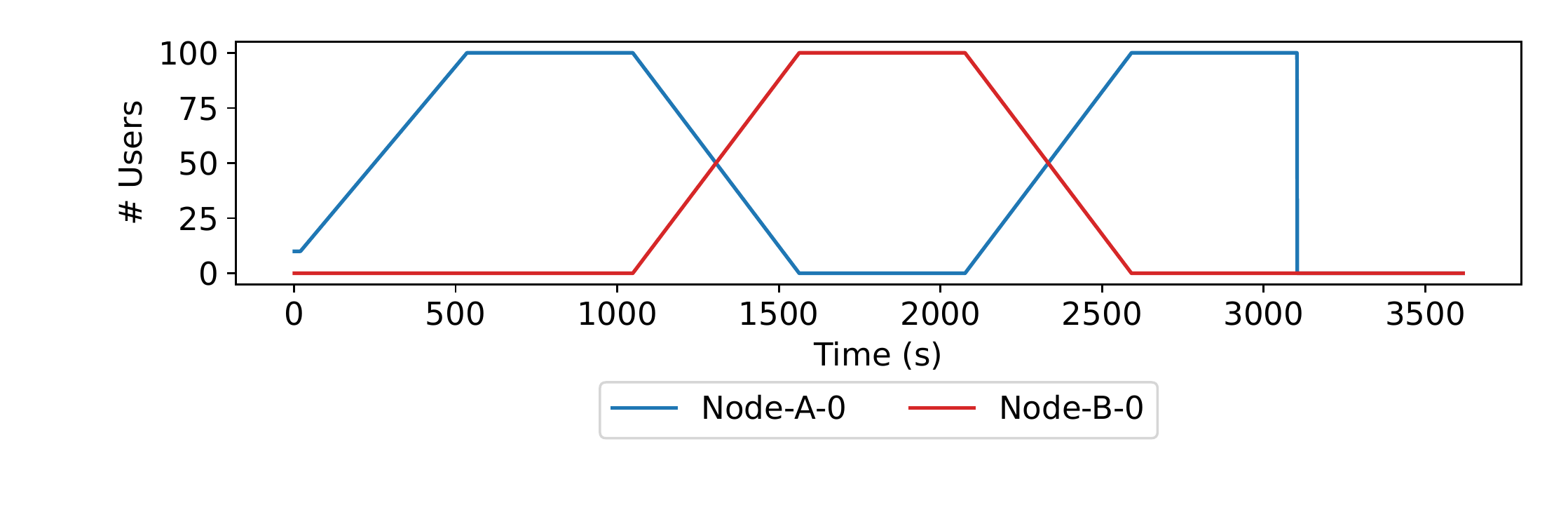}
    \caption{Geo-dynamic workload shape (users)}
    \label{fig:evaluation:migration:wl}
  \end{subfigure}
  \begin{subfigure}{1\columnwidth}
    \centering
    \includegraphics[width=1\columnwidth]{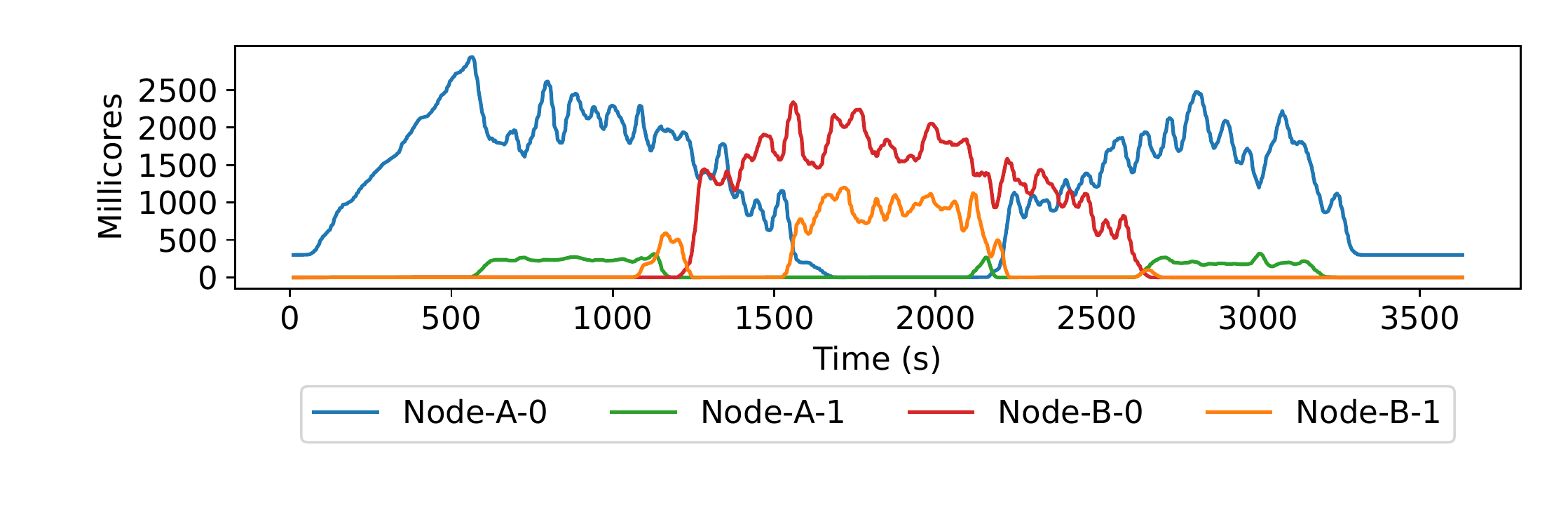}
    \caption{Resource allocation (millicores)}
    \label{fig:evaluation:migration:rs}
  \end{subfigure}
  \begin{subfigure}{1\columnwidth}
    \centering
    \includegraphics[width=1\columnwidth]{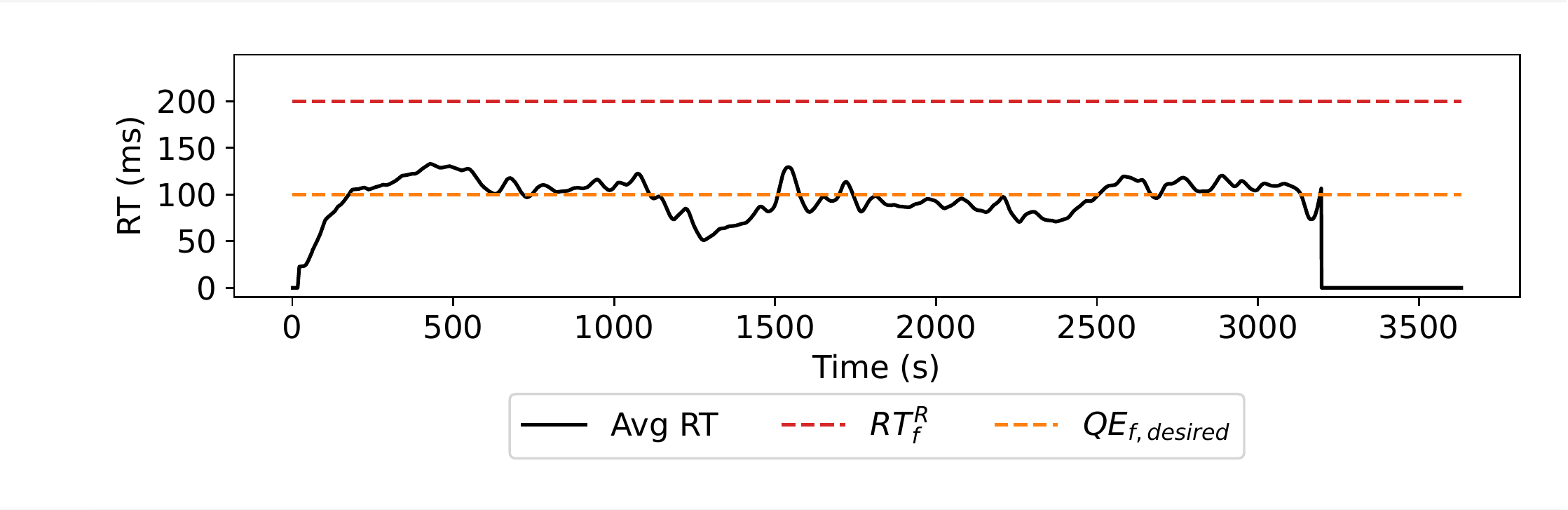}
    \caption{Average response time (ms)}
    \label{fig:evaluation:migration:rt}
  \end{subfigure}
  \begin{subfigure}{1\columnwidth}
    \centering
    \includegraphics[width=1\columnwidth]{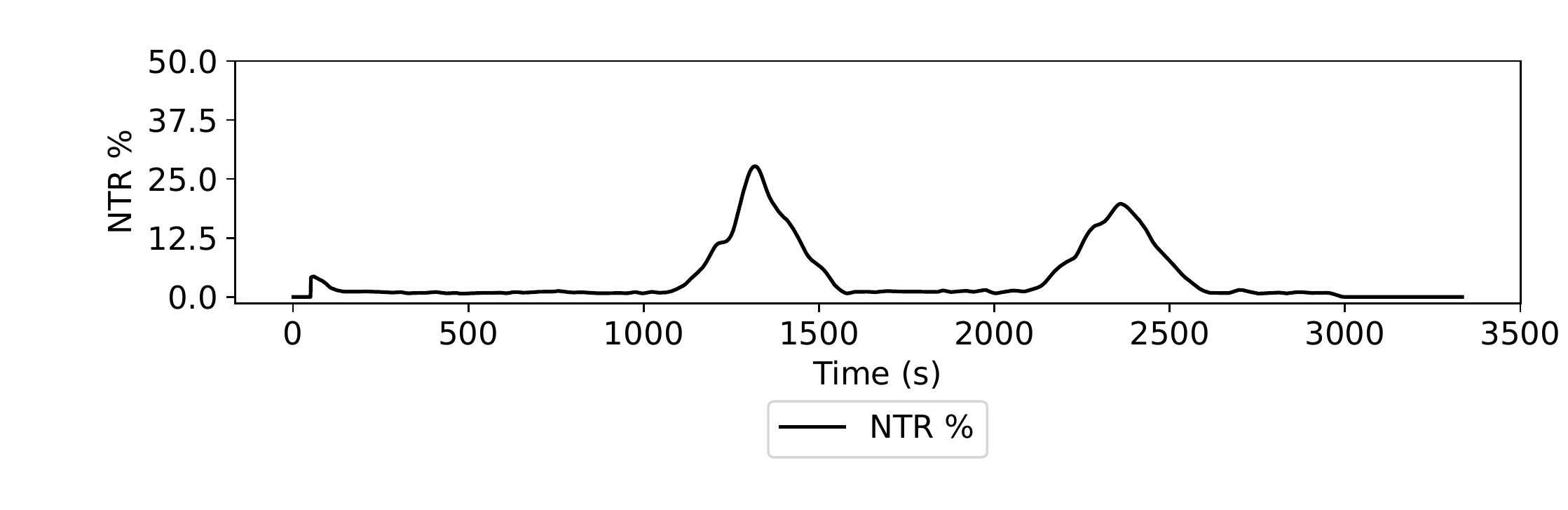}
    \caption{Networking time rate (\%)}
    \label{fig:evaluation:migration:nt}
  \end{subfigure}
  \caption[short]{Behavior of \approach with moving workloads.}
  \label{fig:evaluation:migration}
\end{figure}

Figure~\ref{fig:evaluation:migration} shows the behavior of application \textit{primes} when managed by \approach. Since the multiple runs executed for this set of experiments had similar behavior, the figure illustrates how workloads, resources, and performance varied over time during one of these runs. 
Figure \ref{fig:evaluation:migration:wl} shows how the workload changed in each area. In particular, the workload was generated by users close to node \textit{Node-A-0} for Area A and \textit{Node-B-0} for Area B. Figure \ref{fig:evaluation:migration:rs} presents the resources allocated to each node over time. 
Since communities are independent, at least one instance per function is always allocated (if possible) to minimize cold starts. Thus, the overall allocation is always greater than zero. Conversely, if a node $i$ has $0$ cores allocated at time $t$ for function $f$, it means $f$ is not running on $i$ at $t$ (e.g., from second 0 to ~1250 for Node-B-0). 

The chart shows that if one node in an area cannot manage generated load, the \communitylevel detects this issue and instantiates a new function instance on another node as close to the workload generator as possible. This behavior can be observed close to second $600$ when the workload in Area A reaches the peak and a new replica is created on Node-A-1. Similarly, at second $1500$ a new replica is deployed on Node-B-1 when the workload in Area B increases. In contrast, when the workload decreases instances are deleted as shown close to second $2700$ on Node-B-0.
Moreover, the experiment clearly shows how \approach is able to migrate function instances when users move to keep the network delay minimized. For example, close to second $1000$, users move from Area A to Area B, and right after the function is migrated on node Node-B-0 to handle the workload in the proximity of users.

Thanks to \approach, function \textit{primes} never violates the set response time: the average response time $RT_f$ in Figure \ref{fig:evaluation:migration:rt} is always significantly lower than the threshold (200 ms). The control loops are able to keep the response time very close to the set point.  
Control theoretical controllers behave very well when they operate with high-frequency control loops, enabled by the in-place vertical scaling feature~\cite{DBLP:conf/europar/BaresiQ18,DBLP:conf/icsa/BaresiQ20}.
In fact, the response time only deviates from the set point when the instances are replicated (scaled horizontally), at seconds $600$, $1600$ and $2600$, since the action requires more time than re-configuring containers.
However, note that the response time always returns close to the set point, and this shows that \approach can recover from multiple types of perturbations (e.g, creation and deletion of replicas, fluctuating workloads).

Figure \ref{fig:evaluation:migration:nt} shows that \approach is able to keep the network overhead extremely low. The only peaks in the chart (seconds $1100$ and $2200$) are caused by users who change location and by the fact that routing policies are not updated immediately.

When users start to migrate to another area, replicas cannot always be created immediately on nodes with the minimum network delay, as depicted in the chart close to second $1100$: the workload on Node-B-0 increases and an instance is created on Node-B-1.
This behavior occurs because the two-step optimization process evaluates the placement on B-0 or B-1 to be extremely comparable (within $\epsilon$) since they handle a small portion of the traffic compared to the nodes in Area A.
However, \approach migrates the function instance directly to Node-B-0 node as soon as the workload in Area B increases (close to second $1200$).

\subsection{RQ2: Comparison with other approaches}
\label{sec:evaluation:complex_application}

\begin{table*}[thbp]
\caption[short]{Results of the comparison with other approaches.}
\small
\begin{tabular}{lc|cccc|cccc|cccc|cccc|}
\multicolumn{2}{c|}{\multirow{2}{*}{\textbf{Function}}}                      & \multicolumn{4}{c|}{\textbf{Response time (ms)}}                                                               & \multicolumn{4}{c|}{\textbf{Response time violation (\%)}}                                                     & \multicolumn{4}{c|}{\textbf{Network time rate (\%)}}                                                           & \multicolumn{4}{c|}{\textbf{Core allocation (millicores)}}                                                     \\
                               &     & \multicolumn{1}{c}{NEPT} & \multicolumn{1}{c}{K3S} & \multicolumn{1}{c}{KN} & \multicolumn{1}{c|}{OF} & \multicolumn{1}{c}{NEPT} & \multicolumn{1}{c}{K3S} & \multicolumn{1}{c}{KN} & \multicolumn{1}{c|}{OF} & \multicolumn{1}{c}{NEPT} & \multicolumn{1}{c}{K3S} & \multicolumn{1}{c}{KN} & \multicolumn{1}{c|}{OF} & \multicolumn{1}{c}{NEPT} & \multicolumn{1}{c}{K3S} & \multicolumn{1}{c}{KN} & \multicolumn{1}{c|}{OF} \\ \hline
\multirow{2}{*}{carts-delete}  & $\mu$ & 66.7                     & 64.6                    & 60.6                   & 100.3                   & 0.1                      & 0.3                     & 0                      & 2                       & 3.5                      & 63.9                    & 92.3                   & 72.9                    & 631.3                    & 1921.1                  & 596                    & 597.5                   \\\
                               & $\sigma$ & 3.4                    & 10.9                  & 1.6                  & 27.3                  & 0.1                    & 0.0                   & 0                  & 1.3                   & 2.1                    & 16.5                  & 1.2                  & 18.2                  & 149.9                  & 429.6                 & 2.1                  & 2.6                   \\
\multirow{2}{*}{carts-post}    & $\mu$ & 110.6                    & 175.9                   & 73.8                   & 184.3                   & 0.1                      & 3.5                     & 0.1                    & 3.4                     & 3.7                      & 68                      & 78.3                   & 69                      & 722.8                    & 615.5                   & 597.4                  & 597.3                   \\
                               & $\sigma$ & 7.6                    & 64.2                  & 2.0                  & 73.7                  & 0.1                    & 2.7                   & 0.1                  & 2.6                   & 2.9                    & 22.4                  & 0.9                  & 26.4                  & 178.9                  & 31.3                  & 2.6                  & 2.1                   \\
\multirow{2}{*}{carts-util}         & $\mu$ & 57.4                     & 95.4                    & 54.6                   & 45.6                    & 0                        & 1.7                     & 0                      & 0.1                     & 2.6                      & 78.5                    & 92.8                   & 70.3                    & 516.3                    & 689.3                   & 596.5                  & 4306.1                  \\
                               & $\sigma$ & 3.0                    & 31.0                  & 1.2                  & 1.8                   & 0.1                    & 1.4                   & 0                  &\char`\~ 0                   & 1.2                    & 19.6                  & 1.0                  & 2.2                   & 83.2                   & 162.2                 & 2.2                  & 180.3                 \\
\multirow{2}{*}{catalogue}     & $\mu$ & 53.3                     & 54.6                    & 163.1                  & 39.2                    & 0                        & 0.1                     & 17.7                   & 0                       & 1.6                      & 74                      & 41.9                   & 71.2                    & 102.7                    & 197.6                   & 65.2                   & 458                     \\
                               & $\sigma$ & 2.7                    & 5.2                   & 35.4                 & 1.4                   & 0                    & \char`\~ 0                   & 2.1                  & 0                   & 0.6                    & 11.9                  & 6.3                  & 1.6                   & 4.1                    & 23.9                  & 13.0                 & 3.1                   \\
\multirow{2}{*}{orders}        & $\mu$ & 211.6                    & 418.9                   & 505.1                  & 485.2                   & 0                        & 16.6                    & 16.5                   & 16                      & 4.1                      & 15.8                    & 44.2                   & 25.2                    & 1114.8                   & 4484.5                  & 1040.7                 & 597.8                   \\
                               & $\sigma$ & 12.1                   & 86.7                  & 165.7                & 126.4                 & 0                    & 8.2                   & 3.0                  & 9.0                   & 1.7                    & 7.5                   & 25.0                 & 8.2                   & 273.0                  & 407.2                 & 294.5                & 1.2                   \\
\multirow{2}{*}{payment}       & $\mu$ & 10.4                     & 50.2                    & 27.9                   & 23.6                    & 0                        & 2.8                     & 1.2                    & 0.4                     & 8.2                      & 98.7                    & 98.4                   & 98.9                    & 795                      & 101.8                   & 49.7                   & 443.1                   \\
                               & $\sigma$ & 0.7                    & 9.8                   & 0.4                  & 1.3                   & 0                    & 0.4                   & 0.7                  & \char`\~ 0                   & 6.1                    & 9.5                   & 1.3                  & 3.9                   & 438.9                  & 13.2                  & 0.2                  & 4.3                   \\
\multirow{2}{*}{shipping}      & $\mu$ & 15                       & 75                      & 28.6                   & 88                      & 2.6                      & 5.9                     & 1.5                    & 8.5                     & 6.4                      & 96.2                    & 95.6                   & 92.7                    & 416.5                    & 888.5                   & 597.4                  & 596.9                   \\
                               & $\sigma$ & 1.1                    & 23.2                  & 0.6                  & 32.8                  & 1.1                    & 1.6                   & 0.6                  & 3.5                   & 2.5                    & 16.9                  & 1.7                  & 20.3                  & 132.3                  & 202.8                 & 1.0                  & 0.9                   \\
\multirow{2}{*}{login}    & $\mu$ & 30.3                     & 72.5                    & 73.2                   & 46                      & 0                        & 2.8                     & 11.1                   & 0.2                     & 2.6                      & 70.1                    & 77.9                   & 63.2                    & 76.7                     & 94.2                    & 54.1                   & 452.2                   \\
                               & $\sigma$ & 1.5                    & 12.3                  & 12.1                 & 0.7                   & 0                    & 0.7                   & 7.0                  & \char`\~ 0                   & 0.9                    & 14.4                  & 1.2                  & 1.1                   & 13.9                   & 15.6                  & 6.1                  & 6.6                   \\
\multirow{2}{*}{registration} & $\mu$ & 46.4                     & 57.7                    & 65                     & 34.9                    & 0                        & 0.1                     & 2.6                    & 0                       & 1.4                      & 80.9                    & 87.9                   & 81.7                    & 71.6                     & 105.3                   & 53.6                   & 453.4                   \\
                               & $\sigma$ & 2.7                    & 6.3                   & 4.4                  & 1.3                   & 0                    & 0.1                   & 1.4                  & 0                   & 0.4                    & 10.4                  & 1.6                  & 1.8                   & 9.8                    & 12.0                  & 6.1                  & 6.2                   \\
\multirow{2}{*}{user}          & $\mu$ & 21.8                     & 66.4                    & 177                    & 93.4                    & 0.5                      & 7.8                     & 46.7                   & 16.8                    & 7.1                      & 77.9                    & 31.5                   & 76.8                    & 153.2                    & 681.8                   & 355.3                  & 463.2                   \\
                               & $\sigma$ & 0.7                    & 6.4                   & 35.1                 & 20.3                  & 0.5                    & 0.4                   & 24.2                 & 5.3                   & 1.0                    & 10.1                  & 5.9                  & 16.3                  & 23.3                   & 91.0                  & 166.8                & 1.9                  
\end{tabular}
\label{Tab:evaluation:aggregate}
\end{table*}

We compared our solution against the three approaches described in Section \ref{sec:evaluation:experiment_setup} by means of application \textit{sock-shop}. Note that some of the functions of this application must invoke other functions. For example, function \textit{order} invokes function \textit{user} to retrieve user's address and payment card information, function \textit{catalogue} to retrieve product details, function \textit{payment} to ensure the creation of the invoice and, finally, in case of success, function \textit{carts-delete} to empty the cart. We took these dependencies into account by setting adequate response times as shown in Table ~\ref{Tab:functions}: from 50ms, for simple functions with no dependencies, to 600ms, assigned to the more complex ones.

Each run had a duration of $20$ minutes and used a workload that resembles a steep ramp with an arrival rate $\lambda_{f,i}$ designed to suddenly increase over a short period of time. The workload started with $10$ concurrent users, and we added one additional user every second up to $100$. We considered a network of 6 nodes in Area B and C.

Table~\ref{Tab:evaluation:aggregate}
reports the statistical results obtained during the experiments with each approach and with function of application \textit{sock-shop}.  
The results
show that \approach provided in most of the cases the lowest response time compared to the other approaches. 
The obtained response times were consistent across multiple runs: the standard deviation ranged between $3\%$ and $7\%$ of the average. Other approaches presented higher standard deviation values: in the worst case, KN obtained a standard deviation equal to $32.8\%$ of the average, while K3S ($36.5\%$) and OF ($40\%$) were even more inconsistent.

\approach reported few violations of the required response time. 
For most functions the amount of violations was lower or equal to $0.1\%$, while it was $2.6\%$ and $0.5\%$ for functions \textit{shopping} and  \textit{user}, respectively.
Other solutions obtained significantly higher violations. In the worst case, K3S failed to meet the foreseen response time $16.6\%$ of the requests, while OF and KN reported violations for $16.9\%$ and $46.8\%$, respectively. 
This can be explained because other approaches, compared to \approach,  do not employ precise routing policies, do not perform an adequate resource allocation, and do not solve resource contentions on nodes. 

We can also observe
how \approach routing policies helped meet set response times. The percentage of time spent by routing requests ranges from $1.4\%$ to $8.2\%$ of the total response time, and, on average, only $4.1\%$ of the time is spent in the network. On the other hand, routing policies of other solutions do not consider node utilization, network delay, and applications performance. K3S reported a network time rate ranging from $15.8\%$ to $98.7\%$ of the response time, with an overall average of $72.4\%$. Similarly, OF and KN obtained an average network time rate of $72.2\%$ and $74.1\%$, respectively. 

Finally, 
as for the
resources allocated by each approach for each function, \approach allocated on average $4600$ millicores, while K3S and OF used about twice that amount, $9780$ and $8960$ millicores, respectively. KN uses fewer resources than \approach on average ($4500$ millicores) but it also suffers from a high number of response time violations. This means that KN usually allocates fewer resources than needed (e.g., for function \textit{catalogue}).

Differently from \approach, other solutions do not adopt any resource contention mechanism to provide a fair allocation of resources. For example, K3S allocated most of the resources, $4480$ millicores, to function \textit{orders}, while other functions could not get the resources to work properly. This creates an imbalance among functions that prevents applications to be properly scaled and leads to more response time violations.

\subsection{RQ3: GPU Management}
\label{sec:evaluation:heterogeneous-hw}

The third set of experiments was carried out to assess the transparent GPU management provided by \approach for computationally intensive functions.
To provide a heterogeneous environment, experiments were conducted using the three nodes in Area A (Node-A-2 is equipped with a GPU).

We used two functions, called \textit{resnet-a} and \textit{resnet-b}, both embed the \textit{ResNet} neural network in inference mode. The instances of the two functions deployed on Node-A-2 were set to share the same GPU.

Each run lasted 20 minutes and used the same workload described in Section \ref{sec:evaluation:complex_application} with a number of concurrent users starting from $10$ and up to $30$ (increased by one every second).

\begin{figure}[t]
  \centering
  \begin{subfigure}{1\columnwidth}
    \centering
    \includegraphics[width=\columnwidth]{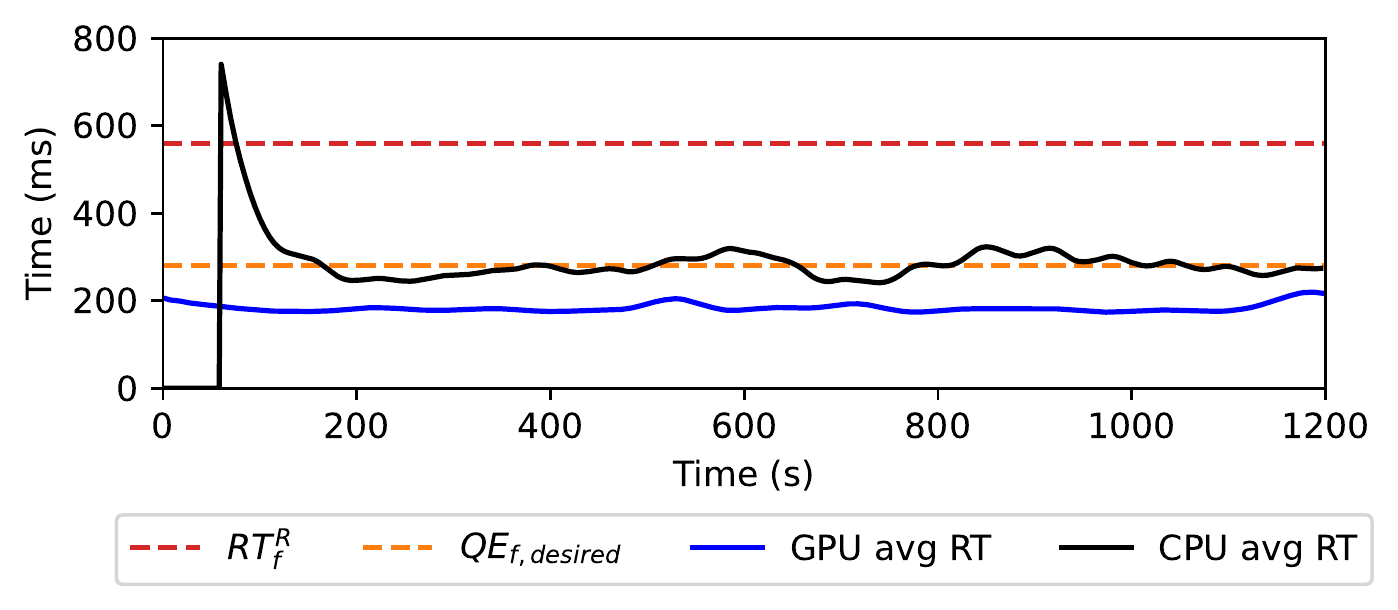}
    \caption{Response time.}
    \label{fig:evaluation:heterogeneous-hw-rt}
  \end{subfigure}
  \begin{subfigure}{1\columnwidth}
    \centering
    \includegraphics[width=\columnwidth]{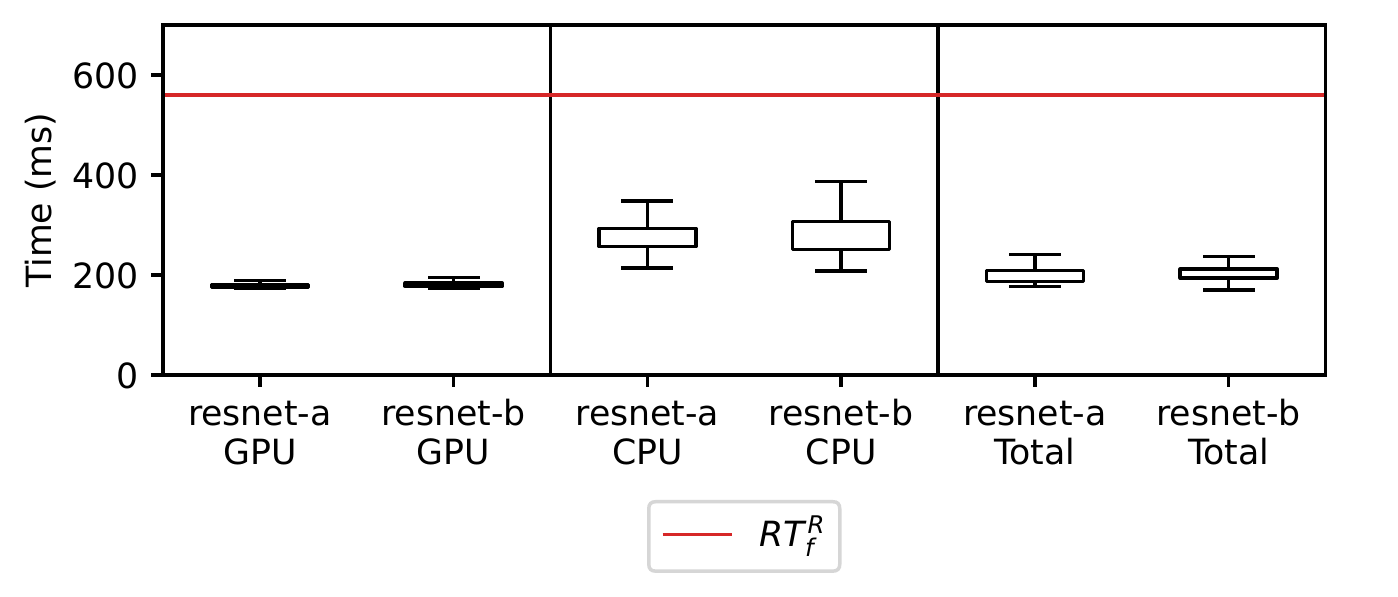}
    \caption{Distributions.}
    \label{fig:evaluation:heterogeneous-hw-box-plot}
  \end{subfigure}
    \caption{Resnet-a: CPU and GPU executions.}
    \label{fig:evaluation:heterogeneous}
\end{figure}

Figure~\ref{fig:evaluation:heterogeneous} illustrates one run of the experiments. 
Figure~\ref{fig:evaluation:heterogeneous-hw-rt} shows the average response time of function resnet-a when executed on both CPUs and GPUs. Function resnet-b obtained similar results that are not reported here for lack of space.
GPU executions obtained an almost constant response time and never violated the set response time. 

At the beginning of the experiment, all the requests were routed to the GPU and after some $50$ seconds the GPU was fully utilized. To avoid degradation of the response time, the \communitylevel quickly reacted by updating the routing policies and allowing part of the workload to be handled by function instances running on CPUs.
The mean response time of CPU instances shows a peak at the beginning of the experiment (with some brief violations of the response time) that is caused by the cold start. After that, the \nodelevel comes into play and dynamically adjusts the CPU cores allocated to the replicas to keep the response time close to the set point. 

The box plot of Figure~\ref{fig:evaluation:heterogeneous-hw-box-plot} shows the distribution of response times for both functions resnet-a and resnet-b on GPU, on CPU, and the aggregated result. The interquartile range (IQR) is set to 1.5, and the rectangle shows the distribution between the 25th and 75th percentiles. 
Both GPU instances of resnet-a and resnet-b are able to keep response times quite far from the set threshold, and thus no violations. In particular, the mean response times of resnet-a and resnet-b are 180ms and 183ms, respectively, which is three times smaller than the threshold.

The distribution of response times on CPUs is wider compared to GPUs. CPU containers are managed by PI controllers that have a transient period to adjust the initial core allocation to an adequate value to reach the desired set point; this does not happen with GPU instances.

Nevertheless, the CPU-only replicas of resnet-a and resnet-b can serve 98.3\% and 100\% of requests within the set response time, respectively.
Moreover, GPU instances handle 70\% of requests while the remaining part was routed to CPU instances. As a result, the total number of violations of both functions is close to $0$.

\subsection{Threats to validity}
\label{sec:evaluation:threats}

We conducted the experiments using twelve functions (three applications) showing that \approach is able to minimize the network delay, to reduce response times, and to efficiently allocate resources compared to other three well-known approaches. However, we must highlight threats that may constrain the validity of obtained results~\cite{DBLP:books/daglib/p/WohlinHH06}:

\textbf{Internal Threats}. The experiments were run with synthetic workloads that may introduce bias. Workloads have a ramp-shape to simulate an incremental growth or reduction of connected users. We used the following procedure to retrieve the maximum concurrent users in each experiment.
First, we fixed the amount and types of nodes the topology was composed of. The maximum concurrent users of each experiment was retrieved by observing how many users were required to generate enough workload to require consistently at least 70\% of the cluster's resources.

The three applications were not provided with a given required response time for each function ($RT^R_f$).
$RT^R_f$ was computed using an iterative process. Starting from $50ms$ and with $50ms$ increments, $RT^R_f$ was set to be able to serve at least 50\% of requests
in an amount of time equal to $RT^R_f/2$.

\textbf{External Threats}. Some of our assumptions may limit the generalization of the experiments. 

Consistently with the serverless paradigm, \approach assumes functions to either be stateless (e.g., without session) or depend on an external database. Currently,
interactions with databases are only partially modeled by \approach.
The time to read from and write on a database is modeled at the \nodelevel as non-controllable stationary disturbance of the response time (e.g., a Gaussian noise).
Thus, during our experiments, databases were deployed on dedicated and properly sized machines. 

Results show that \approach is able to efficiently control functions that depend on a database (e.g., orders, carts-post) with a precision similar to the ones without dependencies (e.g., payment, user). 

\textbf{Construct and Conclusion Threats}. The experiments demonstrate the validity of our claim, that is, that  \approach is able to efficiently execute multiple functions deployed on a distributed edge topology. All experiments have been executed five times and obtained results are statistically robust and show small variance. 

%% file: sections/related-works.tex
\section{Related Work}
\label{sec:related}

The management of edge topologies is a hot and widely addressed topic by both industry and academia~\cite{DBLP:journals/fgcs/BellendorfM20,DBLP:journals/access/JiangCGZW19}. 
To the best of our knowledge, \approach is the first solution that provides: an easy to use serverless interface, optimal function placement and routing policies, in-place vertical scaling of functions, and transparent management of GPUs and CPUs. The relevant related works we are aware of only focus on specific aspects of the problem.

Wang et al.~\cite{DBLP:conf/hpdc/WangAS21} propose \textit{LaSS}, a framework for latency-sensitive edge computations built on top of Kubernetes and Openwhisk\footnote{\url{https://openwhisk.apache.org/}}. LaSS models the problem of resource allocation using a \textit{M/M/c FCFS} queuing model. They provide a fair-share resource allocation algorithm, similar to \approach's \textit{Contention Manager}, and two reclamation policies for freeing allocated resources. 
LaSS is the most similar solution to \approach, but it lacks network overhead minimization and GPU support. Furthermore, the approach is not fully compatible with the Kubernetes API. 
Kubernetes is only used to deploy OpenWhisk. Functions run natively on top of the container runtime (e.g., Docker~\footnote{\url{https://www.docker.com}}) and resources are vertically scaled by bypassing Kubernetes. This approach, also adopted in cloud computing solutions~\cite{DBLP:conf/IEEEcloud/RattihalliGLT19,DBLP:conf/icsa/BaresiQ20}, is known to create state representation inconsistencies between the container runtime and the orchestrator~\cite{DBLP:conf/icsoc/BaresiHQT21}.

Ascigir et al.~\cite{9326369} investigate the problem for serverless functions in hybrid edge-cloud systems and formulate the problem using \textit{Mixed Integer Programming}. They propose both fully-centralized (function orchestration) approaches, where a single controller is in charge of allocating resources, and fully-decentralized (function choreography) ones, where controllers are distributed across the network and decisions are made independently. Compared to \approach, they focus on minimizing the number of unserved requests and they assume that each request can be served in a fixed amount of time (single time-slot). However, this assumption is not easy to ensure in edge computing: nodes may be equipped with different types of hardware and produce different response times. This is naturally considered in \approach with the help of GPUs.

Multiple approaches in the literature focus on placement and routing at the edge~\cite{DBLP:conf/infocom/PoularakisLTTT19,DBLP:conf/infocom/GuoXYY16,DBLP:journals/corr/abs-2108-13222}.
One of the most used techniques, also employed by \approach, is to model the service placement and workload routing as an \textit{Integer} or a \textit{Mixed Integer Programming} problem. 

Notably, Tong et al.~\cite{DBLP:conf/infocom/TongLG16} model a MEC network as a hierarchical tree of geo-distributed servers and formulate the problem as a two-steps \textit{Mixed Nonlinear Integer Programming} (MNIP). In particular, their approach aims to maximize the amount of served requests by means of optimal service placement and resource allocation.
The effectiveness of their approach is verified using formal analysis and large-scale trace-based simulations.
They assume that workloads follow some known stochastic models (Poisson distribution), and that arrival rates are independent and identically distributed. This may not be true in the context of edge computing where workloads are often unpredictable and may significantly deviate from the assumed distribution. \approach does not share these assumptions and uses fast control-theoretical planners to mitigate volatility and unpredictability in the short term.

To cope with dynamic workloads, Tan et al.~\cite{DBLP:conf/infocom/TanHLL17} propose an online algorithm for workload dispatching and scheduling without any assumption about the distribution. However, since their approach only focuses on routing requests, they cannot always minimize network delays, especially when edge clients move from one location to another.

Mobile workloads are addressed, for example, by 
Leyva-Pupo et al.~\cite{DBLP:journals/sensors/Leyva-PupoSC19}, who present a solution based on an \textit{Integer Linear Programming} (ILP) problem with two different objective functions one for mobile users and one for static ones.  Furthermore, since the problem is known to be NP-hard, they use heuristic methods to compute a sub-optimal solution.
Sun et al.~\cite{DBLP:conf/icc/SunA16} propose a service migration solution based on \textit{Mixed Integer Programming} to keep the computation as close as possible to the user. In particular, they consider different factors that contribute to migrations costs (e.g., required time and resources).
However, the two aforementioned solutions exploit virtual machines and they are known for their large image sizes and long start-up times, making service migration a very costly operation. \approach, as other approaches in the literature\cite{DBLP:journals/access/Oleghe21,DBLP:journals/nca/ZhouWWQ20,DBLP:conf/hpdc/WangAS21}, uses containers that are lighter and faster to scale. 

Only a few solutions have been designed for GPU management in the context of edge computing. For example, Subedi et al.~\cite{DBLP:conf/IEEEcloud/SubediHKR21} 
mainly focuses on enabling GPU accelerated edge computation without considering latency-critical aspects such as placing applications close to the edge clients.

%% file: sections/conclusions-future-work.tex
\section{conclusions and future work}\label{sec:conclusions}
\balance
The paper proposes \approach, a serverless-based solution for managing latency-sensitive applications deployed on geo-distributed large scale edge topologies. It provides smart placement and routing to minimize network overhead, dynamic resource allocation to quickly react to workload changes, and transparent management of CPUs and GPUs.
A prototype built on top of K3S, a popular container orchestrator for the edge, helped us demonstrate the feasibility of the approach and interesting results with respect to similar state-of-the-art solutions.

Our future work comprises the improvement of adopted scheduling and resource allocation solutions by exploiting function dependencies~\cite{DBLP:conf/IEEEcloud/LinKW18} and workload predictors to anticipate future demand~\cite{DBLP:journals/fgcs/KumarS18}.
As a further extension, we will consider Bayesian optimization approaches~\cite{DBLP:conf/nips/SnoekLA12,DBLP:conf/recsys/FelicioniDCBHBD20} to find optimal response times automatically. State migration and data consistency approaches can also be integrated to manage stateful applications.
\balance

%% file: sections/acknowledgements.tex
\section{acknowledgements}\label{sec:acknowledgements}
This work has been partially supported by the SISMA national research project (MIUR, PRIN 2017, Contract $201752ENYB$).